\documentclass[aps,prl,floatfix,twocolumn]{revtex4}

\usepackage{graphicx}
\usepackage{epsfig}

\begin{document}

\title{Accuracy of topological entanglement entropy on finite cylinders}

\author{Hong-Chen Jiang$^1$, Rajiv R. P. Singh$^2$, and Leon Balents$^1$}
\affiliation{$^1$Kavli Institute for Theoretical Physics, University
of California, Santa Barbara, CA, 93106\\%
$^2$Physics Department, University of California, Davis, CA, 95616}

\date{\today}

\begin{abstract}%
  Topological phases are unique states of matter which support non-local
  excitations which behave as particles with fractional statistics.  A
  universal characterization of gapped topological phases is provided by
  the topological entanglement entropy (TEE). We study the finite size
  corrections to the TEE by focusing on systems with $Z_2$ topological
  ordered state using density-matrix renormalization group (DMRG) and
  perturbative series expansions.  We find that extrapolations of the
  TEE based on the Renyi entropies with Renyi index $n\geq 2$ suffer
  from much larger finite size corrections than do extrapolations based
  on the von Neumann entropy. In particular, when the circumference of
  the cylinder is about ten times the correlation length, the TEE
  obtained using von Neumann entropy has an error of order $10^{-3}$,
  while for Renyi entropies it can even exceed $40\%$. We discuss the
  relevance of these findings to previous and future searches for
  topological ordered phases, including quantum spin liquids.
\end{abstract}


\maketitle

Topological phases are exotic states of
matter that are characterized by ground state
degeneracy dependent upon global topology of the system on which the
phase resides, and which host exotic excitations with fractional quantum statistics. In
recent years, surprising connections have emerged between
topological phases and quantum information, stimulated by the prospect of
using them to construct an inherently fault-tolerant quantum
computer\cite{Kitaev2003,Nayak2008}. Two dimensional phases with
topological order are well known in connection with the fractional
quantum Hall effect\cite{Laughlin1983}, but are also expected to
exist in frustrated quantum magnets\cite{AndersonRVB,Balents2010}.

While topological phases are not characterized by any {\em local}
order parameter, theory shows that they can be identified by non-local
quantum entanglement, specifically the topological entanglement entropy
(TEE) of the ground states\cite{Kitaev2006,Levin2006}. The entanglement entropy
of a subregion $A$ of the system with a smooth boundary of length $L$
is defined from the reduced density matrix $\rho_A$, according to
$$S_1(A)=-{\rm Tr} [\rho_A {\ln}(\rho_A)],$$ and takes the form
$$S_1(A)=\alpha_1 L-\gamma,$$ where $\gamma$ is the TEE.
Severe finite-size corrections of the formulations in
Ref.\cite{Kitaev2006,Levin2006} due to lattice-scale effects
greatly hinder their practical application\cite{Furukawa2007}.
Instead, two of us have recently proposed a practical and extremely
simple scheme, the ``cylinder construction'', to accurately 
calculate TEE\cite{Jiang2012TEE}. The cylinder construction simply
consists of using the DMRG\cite{White1992DMRG} to calculate the
usual von Neumann entanglement entropy for the division of a
cylinder into two equal halves by a flat cut, and extracting the TEE
from its asymptotic large-circumference limit. Thereby, we can
practically identify topological phases in arbitrary realistic
models, including physical spin
models\cite{Jiang2012TEE,Jiang2011SJ1J2}.

The work above utilized the von Neumann entanglement entropy, and
achieved an accurate extrapolation of the TEE term.   In the
literature, many works study instead the generalized Renyi
entropies, defined as $$S_n(A)=\frac{1}{1-n}\ln {\rm Tr}\rho_A^n,$$
while the von Neumann entropy $S_1$ is defined as the limit
$n\rightarrow 1$. For simplicity, we will call $S_n$ as Renyi
entanglement entropy when $n\geq 2$, while von Neumann entanglement
entropy when $n=1$.   Theoretically, the universal TEE is expected to
obtain also for the Renyi entropy, with $\gamma$ independent of the
Renyi index.  However, extrapolations in the literature based on the
Renyi entropy appear to be substantially less accurate than those
based on the von Neumann entropy, even for larger boundary
lengths $L$\cite{Depenbrock2012Kagome}. In particular, the extrapolated
TEE in Ref.\cite{Depenbrock2012Kagome} from the second-order Renyi
entropy $S_2$ deviates from the expected value with an error an order
of magnitude larger than that from von Neumann entanglement entropy
$S_1$\cite{Jiang2012TEE}. This suggests that extrapolations of Renyi entropies
have significantly larger finite-size effects than von Neumann
entropy.

In this letter, we study the finite-size effects in the TEE
systematically for two canonical models of phases with
$Z_2$ topological order, and
confirm the above suggestion.  We attempt to cast our results in terms
of the expected form,
\begin{eqnarray}
S_n(L)=\alpha_nL-\gamma_n, \label{Eq:TEE}
\end{eqnarray}
as a function of Renyi index $n$.   First, we study the Toric-Code
model whose TEE is known, using DMRG and perturbative series
expansions\cite{BookSeriesExpansion}.   We then turn to the more
realistic $S=1/2$ antiferromagnetic Heisenberg model on the Kagome
lattice. For both cases, we find that the Renyi entropies do have
substantially larger finite-size corrections than the von Neumann
entropy.   We provide some understanding of this tendency from the
fact, which we show from the series expansion, that the line term
$\alpha_n$ varies more rapidly with parameters with increasing $n$.
This makes the extraction of the sub-dominant $\gamma_n$ term less
reliable.

\begin{figure}
\centerline{
    \includegraphics[height=1.8in,width=3.6in] {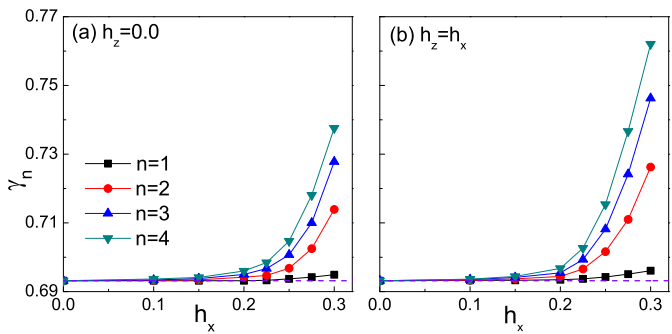}
    }
\caption{(Color online) The extrapolated TEE $\gamma_n$ using von
Neumann entropy (i.e., $n=1$) and Renyi entropy (i.e., $n\geq 2$)
for the Toric-Code model, as a function of the applied magnetic
field $h_x$, for (a)  $h_z=0$ and (b)
symmetric case $h_z=h_x$. The dashed lines represent the expected
universal value, $\ln 2$, for the TEE in the thermodynamic limit.}
\label{Fig:TCHz0SlopeTEEAll}
\end{figure}

\emph{\textbf{Toric-Code Model}} We begin with the well known Toric-Code
model (TCM)\cite{Kitaev2003}, (See
Supplementary Information for notational details). Without
applied fields, i.e., $h_x=h_z=0$, the pure TCM is exactly
soluble\cite{Kitaev2003} with a $Z_2$ topological ordered ground
state. After turning on the magnetic fields, the model is no longer
exactly soluble. Previous studies
\cite{Trebst2007,Vidal2008,Tupitsyn2008} show that the $Z_2$
topological phase remains stable and robust until the
magnetic fields are large enough that the system undergoes a
phase transition from the topological phase to a trivial one.
Numerically, Jiang, Wang and Balents have systematically calculated
the von Neumann entanglement entropy $S_1$ using cylinder
construction\cite{Jiang2012TEE}, and extrapolated an accurate TEE
$\gamma_1=\ln(2)$ in the $Z_2$ topological phase, even very close to
the phase transition point. In this paper, we further calculate the
Renyi entropy to study the finite-size effects on the TEE $\gamma_n$
as a function of Renyi index $n$. To make sure that $S_n$ only
scales with the cylinder circumference $L_y=L$, we will work in the
long cylinder limit, i.e., cylinder length $L_x\rightarrow \infty$.
Therefore, we can directly extrapolate the TEE $\gamma_n$ using
Eq.(\ref{Eq:TEE}).

\begin{figure}
\centerline{
    \includegraphics[height=1.8in,width=3.6in] {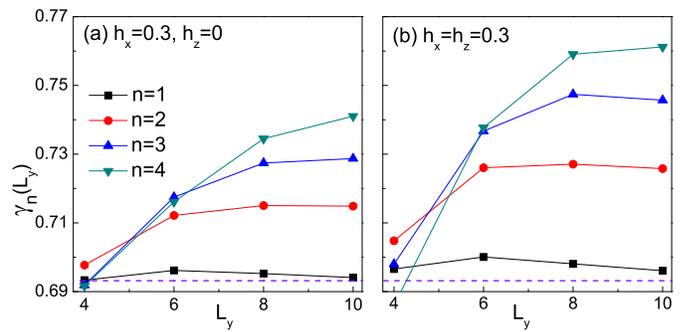}
    }
\caption{(Color online) Moving two point data fits for the
extrapolated TEE $\gamma_n$ for the Toric-Code model for (a) $h_x=0.3$ with $h_z=0$, and (b)
$h_x=h_z=0.3$, as a function of the
cylinder circumference $L_y$ and entropy index $n$.   Here,
$\gamma_n(L_y)$ are fitted using two data points, i.e., $S_n(L_y)$
and $S_n(L_y+2)$ using Eq.(\ref{Eq:TEE}). The dashed lines represent the
expected asymptotic value $\gamma =\ln 2$. } \label{Fig:TCHz0SlopeTEE}
\end{figure}

\begin{figure}
\centerline{
    \includegraphics[height=2.6in,width=3.4in] {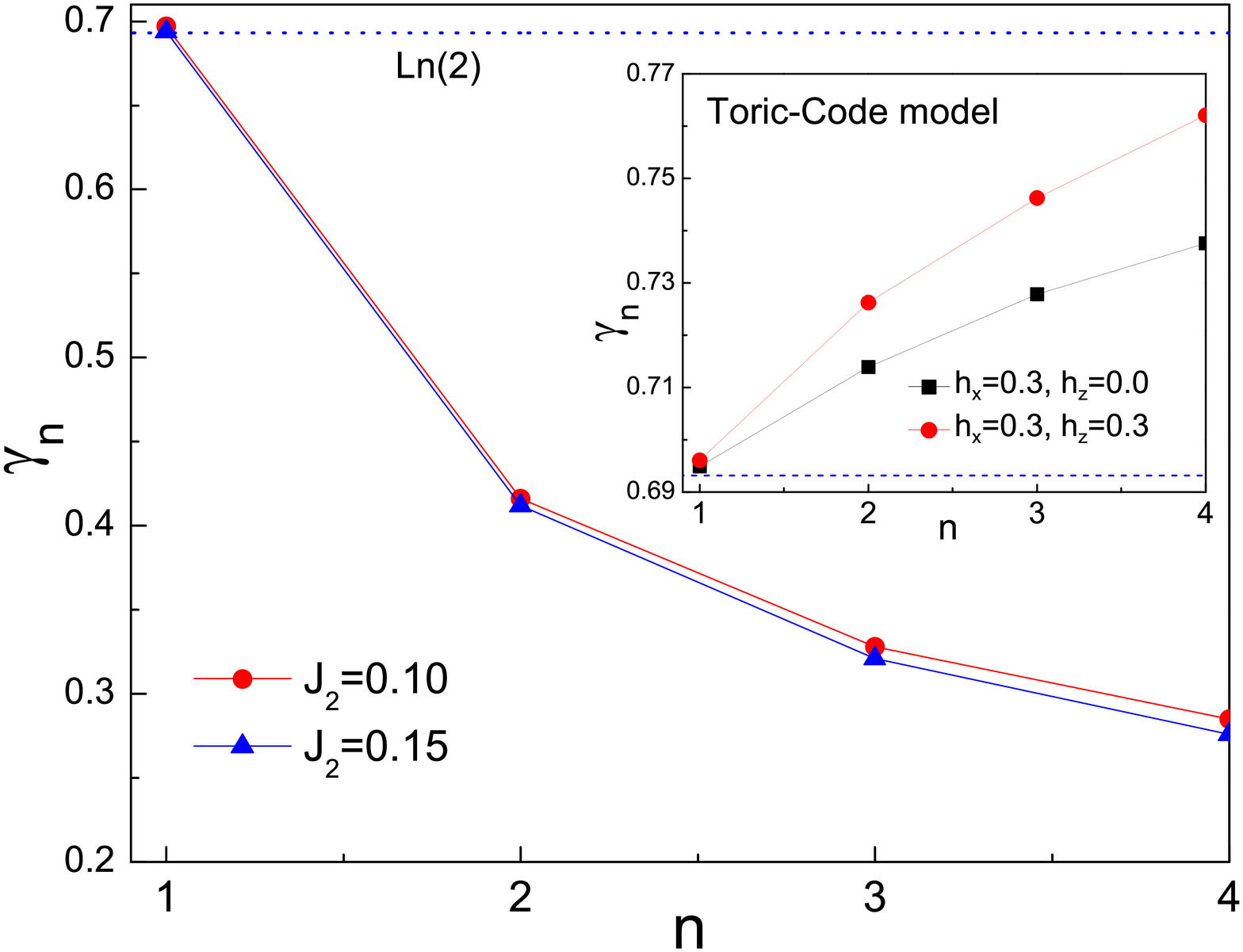}
    }
\caption{(Color online) The extrapolated TEE $\gamma_n$ for the Kagome
  J$_1$-J$_2$ Heisenberg model
as a function of entropy index $n$ at both $J_2=0.10$ and
$J_2=0.15$. Inset: $\gamma_n$ for TCM as a function of entropy index
$n$, for the  case $h_x=0.3, h_z=0$ (black square), and the
symmetric case $h_x=h_z=0.3$ (red circle). The dashed lines
correspond to $\ln(2)$.} \label{Fig:KagomeTEE}
\end{figure}

To see the finite-size corrections, we first consider an applied
magnetic field only along $x$-direction, i.e., $h_z=0$. The
extrapolated TEE $\gamma_n$ are shown in
Fig.\ref{Fig:TCHz0SlopeTEEAll} (a) as a function of magnetic field
$h_x$ for fields within the topological phase.  As shown in the
Supplementary Material, for the fields in
Fig.~\ref{Fig:TCHz0SlopeTEEAll}, the spin correlation length remains
of order one lattice spacing or smaller.  We see that the linear fit
using data for $L_y=6\sim 12$ and Eq.(\ref{Eq:TEE}) gives quite
accurate results for $\gamma_1$ for $h_x<h_x^c\approx
0.328$\cite{Trebst2007}. Even for $h_x=0.3$, very close to the
quantum phase transition, we obtain $\gamma_1=0.6945(20)$, which is
accurate to the expected value (i.e., the dashed line) to a fraction
of percent $\sim 0.2\%$.  By contrast, the estimated TEE $\gamma_n$
obtained from the Renyi entropy for $n\geq 2$ shows dramatically
larger deviations from the universal value.  These deviations grow
when approaching the phase transition point, where the errors become
an order of magnitude larger than those for $\gamma_1$. For example,
$\gamma_2=0.714(5)$ at $h_x=0.3$, with an error around $\sim 3\%$.
The deviations also increase with Renyi index $n$, e.g., $\sim 7\%$
for $n=4$, as shown in Fig.\ref{Fig:TCHz0SlopeTEEAll} (a) and inset
of Fig.\ref{Fig:KagomeTEE}. Similar results hold for the symmetric
case $h_x=h_z$, with even larger finite-size corrections, as shown
in Fig.\ref{Fig:TCHz0SlopeTEEAll}(b) and inset of
Fig.\ref{Fig:KagomeTEE}.   Systematically,  the finite-size
corrections to the TEE $\gamma_n$ defined by Eq.(\ref{Eq:TEE}) are
much larger for the Renyi entropy than for the von Neumann entropy.

Our expectation is that {\em in the thermodynamic limit} $L_y
\rightarrow \infty$, all $\gamma_n$ should converge properly to the
universal value.  We look for signs of this tendency, by using a
moving two data-point fit. We can define $\gamma_n(L_y)$ using two
data points with different cylinder circumferences $L_y$ and
$L_y+2$. Examples of such $\gamma_n(L_y)$ are shown in
Fig.\ref{Fig:TCHz0SlopeTEE} for $h_x=0.3$ and $h_z=0$ in (a), and
for $h_z=h_x=0.3$ in (b), as a function of $L_y$. For both cases,
$\gamma_1(L_y)$ quickly converges to the expected value, which is in
sharp contrast to $\gamma_n(L_y)$ using Renyi entropies with $n\geq
2$.  For the latter cases, the TEE is systematically over-estimated.
The curves in Fig.~\ref{Fig:TCHz0SlopeTEE} do at least show downward
curvature, consistent with eventual convergence to the universal
value for larger $L_y$, but it is clear the much larger systems
would be required to test this in detail.

To understand the origin of the $n$-dependence in the extrapolations, we
turn to a perturbative series expansion calculation of the line term
$\alpha_n$. Since the TEE is obtained after subtraction of the much
larger line term, an accurate extrapolation of $\gamma_n$ also requires an
accurate calculation of $\alpha_n$.  While there have been perturbative
studies of the Kitaev model in a field\cite{kpschmidt} and some
entanglement properties have also been studied\cite{fradkin,hamma}, we
are not aware of any exact perturbative evaluation of its line entropy
relevant to our geometry. In order to carry out this calculation, we
turn to Linked Cluster Methods.\cite{BookSeriesExpansion}

We consider one of the ground state sectors of the TCM and
map the problem on to the Transverse-Field Ising Model (TFIM) if only
$h_x$ is non-zero\cite{Trebst2007} and on to a $Z_2$ Lattice-gauge model
if both $h_x$ and $h_z$ are non-zero.\cite{Tupitsyn2008} In either case,
one has a unique non-degenerate ground state.  Here, we will consider
only the former mapping and the case with only $h_x$ non-zero. Up to
$4$-th order, the dependence on $h_x$ and $h_z$ are additive and hence
knowing the dependence on $h_x$ and the symmetry under the interchange
of $h_x$ and $h_z$ one can easily write down the full dependence on
$h_x$ and $h_z$.

Given the exact mapping between the models, if we were to calculate
some property like the ground state energy in a series expansion in
the field, the expansions would be identical to the TFIM. However,
there is a crucial difference for the entanglement
entropies.\cite{ann} The TFIM variables sit at the center of the
plaquettes whereas the TCM variables and the perturbing fields live
on the bonds. Furthermore, each state in the TCM is a linear
superposition of many basis states (say in the $\sigma_z$ basis).
This affects how the states are represented on either side of the
partition and hence the reduced density matrix as we discuss below.

\begin{figure}
\centerline{
    \includegraphics[height=2.4in,width=3.2in] {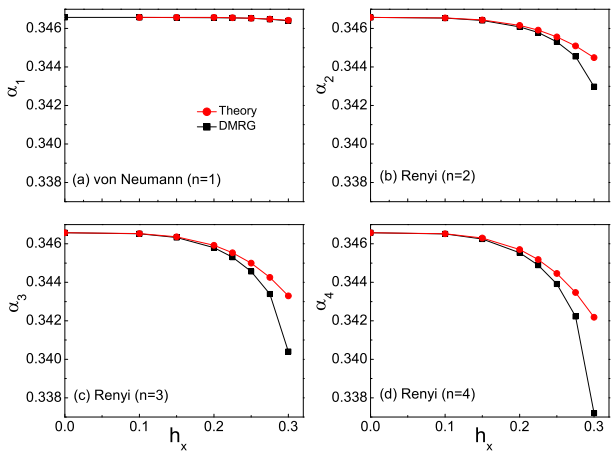}
    }
\caption{(Color online) Line entropy $\alpha_n$ obtained from DMRG
(block square) and perturbative calculation (red circle) for both
von Neumann entropy (i.e., $n=1$) and Renyi entropy (i.e., $n\geq
2$) for the Toric-Code model, as a function of the magnetic field
$h_x$, here $h_z=0$.} \label{Fig:TCHz0Slope}
\end{figure}

In the linked cluster method, we can define a cluster by a set of bonds, where
the perturbative fields are present.\cite{oitmaa,gelfand-jsp}
The entanglement entropies can be expressed as
\begin{equation}
S_n = \sum_c W_n(c),
\end{equation}
where the sum is over all possible clusters $c$
and the weight of a cluster $W_n(c)$ is defined recursively
by the relations
\begin{equation}
W_n(c)=S_n(c)-\sum_{s} W_n(s),
\end{equation}
where $S_n(c)$ is the entanglement entropy for the cluster $c$ and
the sum over $s$ is over all subclusters of $c$.

One can show that, for the line entropy, the only clusters that will
give non-zero contributions in powers of $h$ are those that are (i)
linked and (ii) that contain at least one bond in subsystem A and
one in subsystem B.\cite{oitmaa,gelfand-jsp} Here, one should note
that two bonds are linked if they meet at a site or if they are on
the opposite sides of an elementary plaquette (because they can both
change the flux through a common plaquette). This implies that all
such clusters must be situated close to the interface between A and
B. Since such linked clusters can be translated along the line, it
follows that, for a large system, this entropy is proportional to
the length of the line. In fourth order, we can group the
perturbations into just two distinct clusters, whose calculational
details can be found in the Supplementary materials.

For line term of the Renyi entropy for $n>1$, to order $h^4$, we obtain
\begin{equation}
\alpha_n = {1\over 2}(\ln{2} - {9n\over 32}h^4 +{3n\over n-1} {h^4\over 128})
\label{Eq:SESlopeAlphaRN}
\end{equation}
while, the von-Neumann entropy ($n\to 1$) becomes
\begin{equation}
\alpha_1 = {1\over 2}(\ln{2} - ({33\over 128}-{5\over 32} \ln{2})h^4 -{3\over
32} h^4\ln{h}). \label{Eq:SESlopeAlphaVN}
\end{equation}
Note that the innocuous $h^n\ln{h}$ singularity is inevitable for
the von-Neumann entropy in any model, where there are Schmidt
states, whose weight is zero in the unperturbed model but becomes
non-zero as a power of the perturbation parameter.

In. Fig.\ref{Fig:TCHz0Slope} we show the line entropies $\alpha_n$
obtained in DMRG compared with the series expansion results (See
also Fig.\ref{FigS:TCHz0dSlope} in Supplementary materials). The
agreement is excellent. The important thing to note is the $n$
dependence of the line entropy. The linear $n$ dependence in
Eq.(\ref{Eq:SESlopeAlphaRN}) means that with increasing $n$, the
line entropy changes more rapidly with the applied fields. Since,
the topological entanglement entropy is obtained after subtraction
of the much larger line entropy, it follows that with increasing $n$
the entanglement entropy would have a much larger finite size
correction, as the correlation length increases.

\emph{\textbf{Kagome Heisenberg Model}} We now turn to the
spin-$1/2$ Heisenberg model on the Kagome lattice, for which
compelling\cite{Jiang2008Kagome,White2011Kagome} and direct
evidence\cite{Depenbrock2012Kagome,Jiang2012TEE} for a $Z_2$
topological quantum spin liquid has been obtained by extensive DMRG
studies. In particular, highly accurate TEE $\gamma_1=\ln(2)$ has
been obtained using cylinder construction\cite{Jiang2012TEE}, for
the model with both first- and second-neighbor interactions -- see
the Supplementary Information for details of the definition.
Specifically, the extrapolated TEE $\gamma_{1}=0.698(8)$ at
$J_2=0.10$ and $\gamma_{1}=0.694(6)$ at $J_2=0.15$, both within one
percent of $\ln(2)=0.693$.

As in the TCM, we find that the Renyi entropies give much less accurate
estimates of $\gamma_n$.  The extrapolated TEE $\gamma_n$ for $n\geq 2$
clearly deviates from the expected value, even when the cylinder
circumference is much larger than the correlation length, i.e.,
$L_y\approx 10 \xi$ (the correlation lengths are known from the earlier
study in Ref.\cite{Jiang2012TEE}). For example, as shown in
Fig.\ref{FigS:KagomeHeisenbergModelSN}, a linear fit using
Eq.(\ref{Eq:TEE}) gives $\gamma_2\approx 0.42(1)$ at both $J_2=0.10$ and
$J_2=0.15$: a huge error of $\sim 40\%$.   Moreover, with increasing Renyi
index $n$, the deviation becomes even larger, reaching, for example $\sim 60\%$ for
$n=4$, as shown in Fig.\ref{Fig:KagomeTEE}.  These results show that
large finite-size corrections to the Renyi entropies obtain not only in
the ``artificial'' TCM, but also in realistic quantum spin
Hamiltonians.

\emph{\textbf{Summary and Conclusion}} In this paper, we studied the
finite-size scaling of the TEE for systems with $Z_2$ topological order,
using DMRG simulations and perturbative series expansions. We find that
generally the finite-size errors in the TEE based on the Renyi entropy
estimators (i.e., $n\geq 2$) are much larger than those obtained from
the von Neumann entropy (i.e., $n=1$). In particular, when the cylinder
circumference is around ten times the correlation length, $L_y\approx
10\xi$, the extrapolated TEE using von Neumann entropy is quite accurate
with an error of order $10^{-3}$.  On the contrary, the error can be
orders of magnitude larger for Renyi entropy. For instance, for
spin-$1/2$ Kagome Heisenberg model, the error is only around a fraction
of percent for von Neumann entropy, while it is $40\%$ or even larger
for Renyi entropy.  Perturbative study of the TCM shows that the larger
finite size corrections for the TEE originates in part from the enhanced
variation with parameters of the line entropy $\alpha_n$ with increasing
$n$.  This indicates moreover that estimates of the TEE become less
accurate with increasing $n$.  We note that errors of the magnitude
found here for $n\geq 2$ in the Kagome Heisenberg model are large enough
to perhaps preclude a definitive identification of the topological
phase, even if we {\em assume} that a universal value obtains in the
thermodynamic limit.  Our results clearly indicate that great care must
be taken into account for finite size corrections in numerical
calculations of the TEE, particularly those based on Renyi entropies
with $n\geq 2$.  This gives techniques, such as DMRG, which have direct
access to the full density matrix and hence von Neumann entropy, a
distinct advantage.  Although in this paper we have focused on systems with
$Z_2$ topological order, similar conclusions may be expected more
generally.

\emph{\textbf{Acknowledgement}}  HCJ thanks Andreas Ludwig for
helpful discussion and Zhenghan Wang for early collaboration. We
acknowledge computing support from the Center for Scientific
Computing at the CNSI and MRL: an NSF MRSEC (DMR-1121053) and NSF
CNS-0960316.   This research was supported in part by DMR-1004231
(RRPS) and DMR-1206809 (LB) and the KITP NSF grant PHY-1125915 (HCJ,
LB, RRPS).


\appendix

\begin{center}
\noindent {\large {\bf Supplementary Information}}
\end{center}



\renewcommand{\thefigure}{S\arabic{figure}}
\setcounter{figure}{0}
\renewcommand{\theequation}{S\arabic{equation}}
\setcounter{equation}{0}


\subsection{I. Toric-Code Model in Magnetic Field}%
The Toric-Code model\cite{Kitaev2003} with an applied magnetic field
is given by
\begin{eqnarray}
H=-J_s\sum_s A_s - J_p\sum_p B_p - h_x\sum_i\sigma^x_i -
h_z\sum_i\sigma^z_i,\label{EqS:ToricCodeModel}
\end{eqnarray}
where $\sigma^x_i$ and $\sigma^z_i$ are Pauli matrices, $A_s=\Pi_{i\in
  s}\sigma^x_i$ and $B_p=\Pi_{i\in p}\sigma^z_i$.  Subscripts $s$ and
$p$ refer respectively to vertices and plaquettes on the square lattice,
whereas $i$ runs over all bonds where spin degrees of freedom are
located. Without magnetic field, i.e., $h_x=h_z=0$, the pure TCM can be
solved exactly\cite{Kitaev2003}.  It exhibits a 4-fold ground state
degeneracy on the torus, and has $Z_2$ topological order with total
quantum dimension $D=2$. All elementary excitations are gapped and
characterized by eigenvalues $A_s=-1$ (a $Z_2$ charge on site $s$) and
$B_p=-1$ (a $Z_2$ vortex on plaquette $p$). After turning on the
magnetic fields, the model cannot be solved exactly anymore. However,
previous studies\cite{Trebst2007,Vidal2008,Tupitsyn2008} show that the
$Z_2$ topological phase remains quite stable and robust until the
magnetic fields are large enough to induce a phase transition from the
topological phase to the trivial one.  Specifically, such a phase
transition takes place at the critical magnetic field $h_x^c\approx
0.32$ when $h_z=0$, while $h_x^c\approx 0.34$ along the symmetric line
with $h_z=h_x$.

\begin{figure}
\centerline{\includegraphics[height=2.4in,width=3.2in]{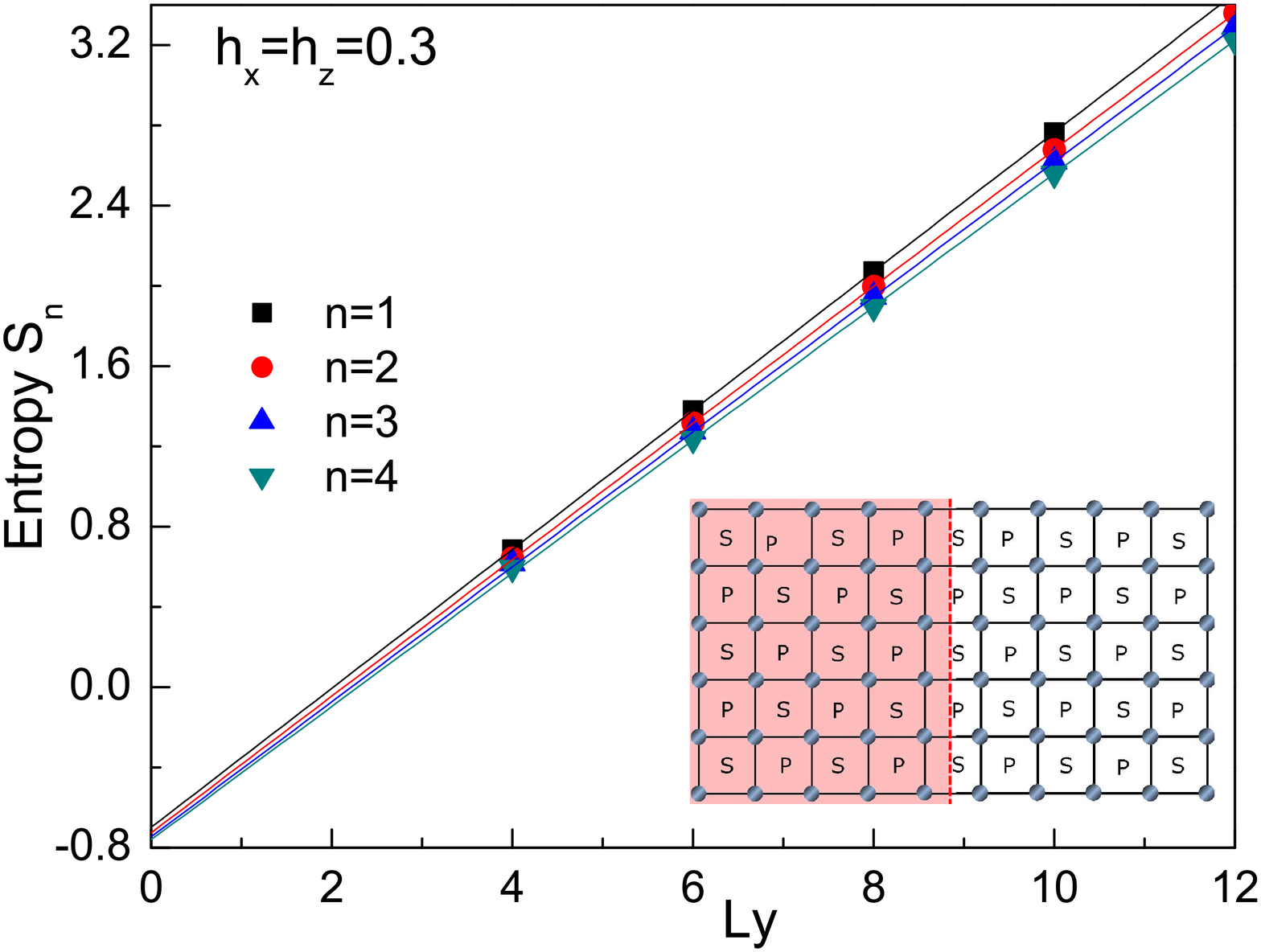}}%
\caption{The entanglement entropy for the Toric-Code model in
magnetic fields $h_x=h_z=0.3$ in Eq.(\ref{EqS:ToricCodeModel}), with
$L_y=4\sim 12$ at $L_x=\infty$. Here $n$ is the entropy index.
Inset: Square lattice with $L_x=10$ and $L_y=6$. Here $S$ represents
the star operator $A_s$, while $P$ represents the plaquette operator
$B_p$.} \label{FigS:ToricCodeModel}
\end{figure}

For the DMRG simulation, we consider an equivalent square lattice,
where the spin operators $\sigma^x$ and $\sigma^z$ sit on the sites
instead of the bonds. Therefore, the star operator $A_s$ and the
plaquette operator $B_p$ of the original lattice now sit on
alternating plaquettes in the equivalent square lattice, as shown in
Figure~\ref{EqS:ToricCodeModel}, labeled as $S$ and $P$,
respectively. Note that on this equivalent square lattice, there are
an even number of dangling spins within each plaquette at the open
edges. For the pure Toric-Code model with cylinder boundary
condition, the first $2^{L_y/2-1}$ eigenvalues of the reduced
density matrix $\rho_A$ are degenerate and equal to $1/2^{L_y/2-1}$,
while all the other eigenvalues are zero. Therefore, the
entanglement entropy, defined as $S_n=\frac{1}{n-1}\ln {\rm
Tr}\rho^n_A$, does not depend on the entropy index $n$, and is equal
to $S_n=\frac{\ln(2)}{2}L_y-\ln(2)$, with coefficient
$\alpha_n=\ln(2)/2$ and TEE $\gamma_n=\ln(2)$.

\begin{figure}
\centerline{\includegraphics[height=2.4in,width=3.2in]{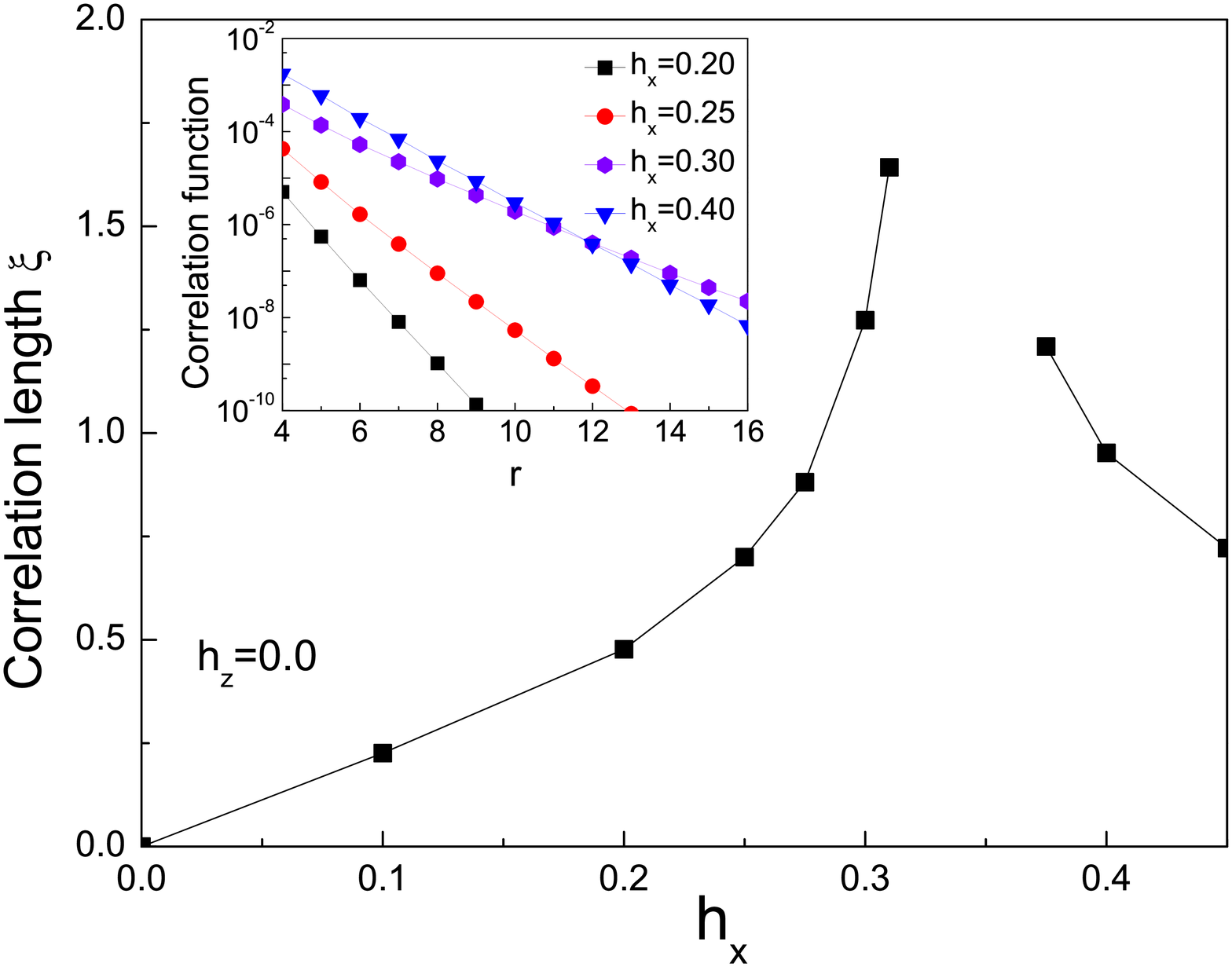}}%
\caption{The spin-spin correlation length for the Toric-Code model
as a function of the applied magnetic field $h_x$ in
Eq.(\ref{EqS:ToricCodeModel}), with $h_z=0.0$. Inset: Examples of
spin-spin correlation function $\langle \tilde{\sigma}^x_{0}
\tilde{\sigma}^x_{r}\rangle=\langle \sigma^x_{0}
\sigma^x_{r}\rangle-\langle \sigma^x_{0}\rangle
\langle\sigma^x_{r}\rangle$ along the cylinder, at different
magnetic field $h_x$.} \label{FigS:ToricCodeModelCorLen}
\end{figure}

\begin{figure}
\centerline{
    \includegraphics[height=4.4in,width=3.0in] {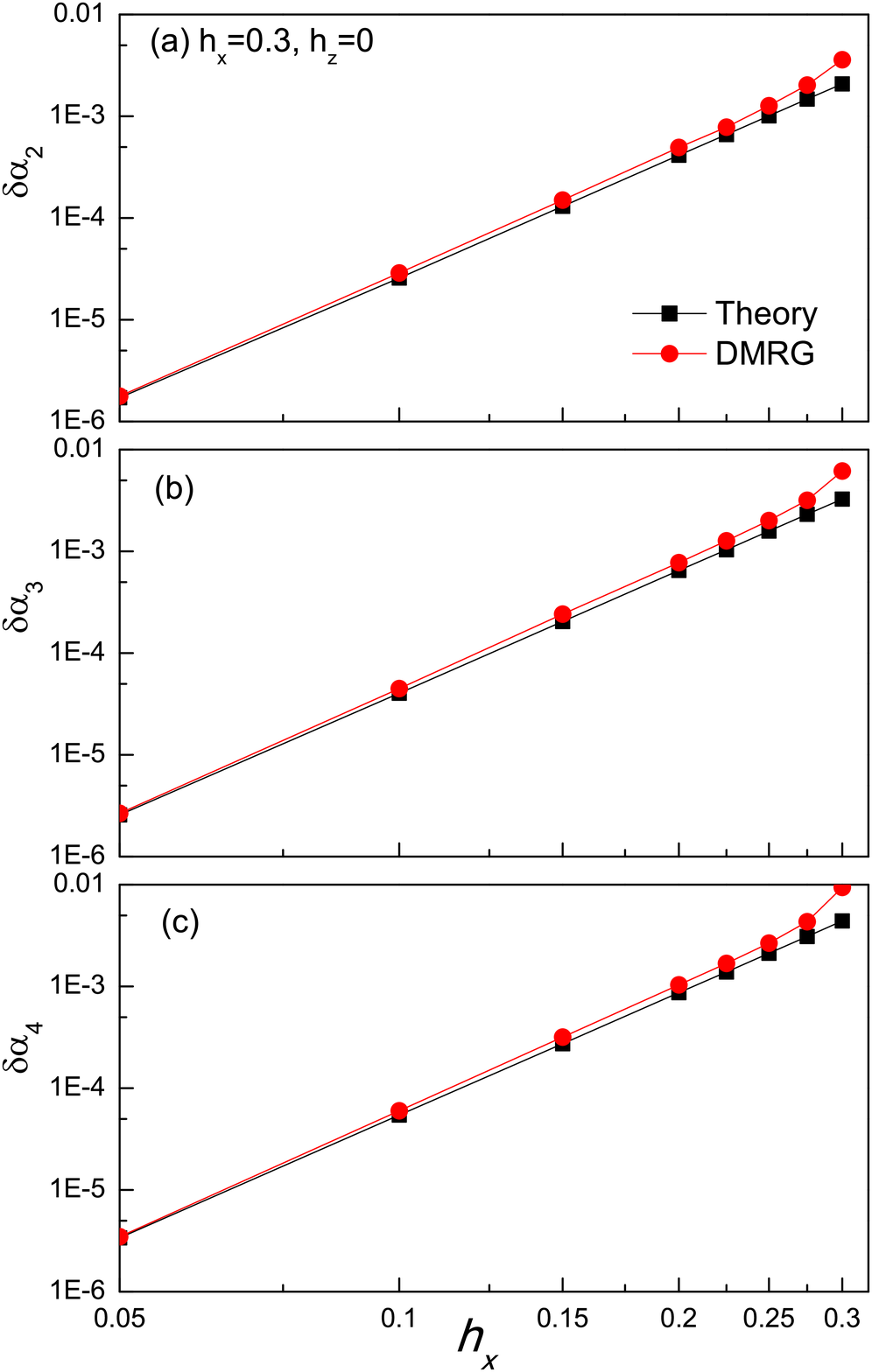}
    }
\caption{Deviation of the coefficient $\delta \alpha_n\equiv
\alpha_n(h_x=0)-\alpha_n(h_x)$ obtained from DMRG using
Eq.(\ref{Eq:TEE}) (red circle) and theoretical formula in
Eq.(\ref{Eq:SESlopeAlphaRN}) up to 4th order series expansion for
Renyi entropy (black square) with (a) $n=2$, (b) $n=3$ and (c) $n=4$
for the Toric-Code model, as a function of the applied magnetic
field $h_x$. Here $h_z=0$.} \label{FigS:TCHz0dSlope}
\end{figure}

After turning on the magnetic field, the degeneracy of entanglement
spectrum is lifted, and the correlation length $\xi$ becomes finite, as
shown in Fig.\ref{FigS:ToricCodeModelCorLen} (note that other types of
the correlation length may be defined, but are expected to differ just
by a factor of order 1 from the above one) . Therefore, the entanglement
entropy $S_n$ will depend on the entropy index $n$, and it follows that
so too will the coefficient $\alpha_n$ and the extrapolated TEE
$\gamma_n$. Indeed, our results show that when approaching the phase
transition point (e.g., $h_x^c=0.34$ along the symmetric line with
$h_x=h_z$), the dependence of the TEE $\gamma_n$ on $n$ becomes clearer
and clearer, as shown in Fig.\ref{Fig:TCHz0SlopeTEEAll}(b) in the main
text, and Fig.\ref{EqS:ToricCodeModel} here. For example, the
correlation length $\xi\approx 1$ lattice spacing at $h_x^c=0.30$, and the
resulting TEE $\gamma_1=0.696(4)$ obtained from the von Neumann entanglement entropy
is still quite accurate with an error around a few fraction of
$10^{-3}$, i.e., $\sim 0.4\%$.   On the contrary, the fitted TEE
$\gamma_n$ using Renyi entropy (i.e., $n\geq 2$) deviates from the
expected value clearly, with an error more than order of magnitude
larger than for $\gamma_1$. For instance, $\gamma_2=0.726(6)$, with an error
around $5\%$, around one order of magnitude larger than that for
$\gamma_1$. Such a deviation is even larger with the increase of $n$,
e.g., $\gamma_4=0.762(8)$, with an error around $10\%$, more than one
order of magnitude larger than that for $\gamma_1$.

Deviations are also found in the line entropies $\alpha_n$.   Fig.\ref{FigS:TCHz0dSlope} shows the
comparison for $\alpha_n$ obtained numerically using
Eq.(\ref{Eq:TEE}) and theoretically using
Eq.(\ref{Eq:SESlopeAlphaRN}) up to 4th order series expansion.   The
series expansion results match the DMRG values remarkably accurately,
though as expected, the difference between the two increases markedly on
approaching the phase transition point, especially for larger $n$.

\subsection{II. Perturbative calculations}

\begin{figure}
\begin{center}
 \includegraphics[width=6cm]{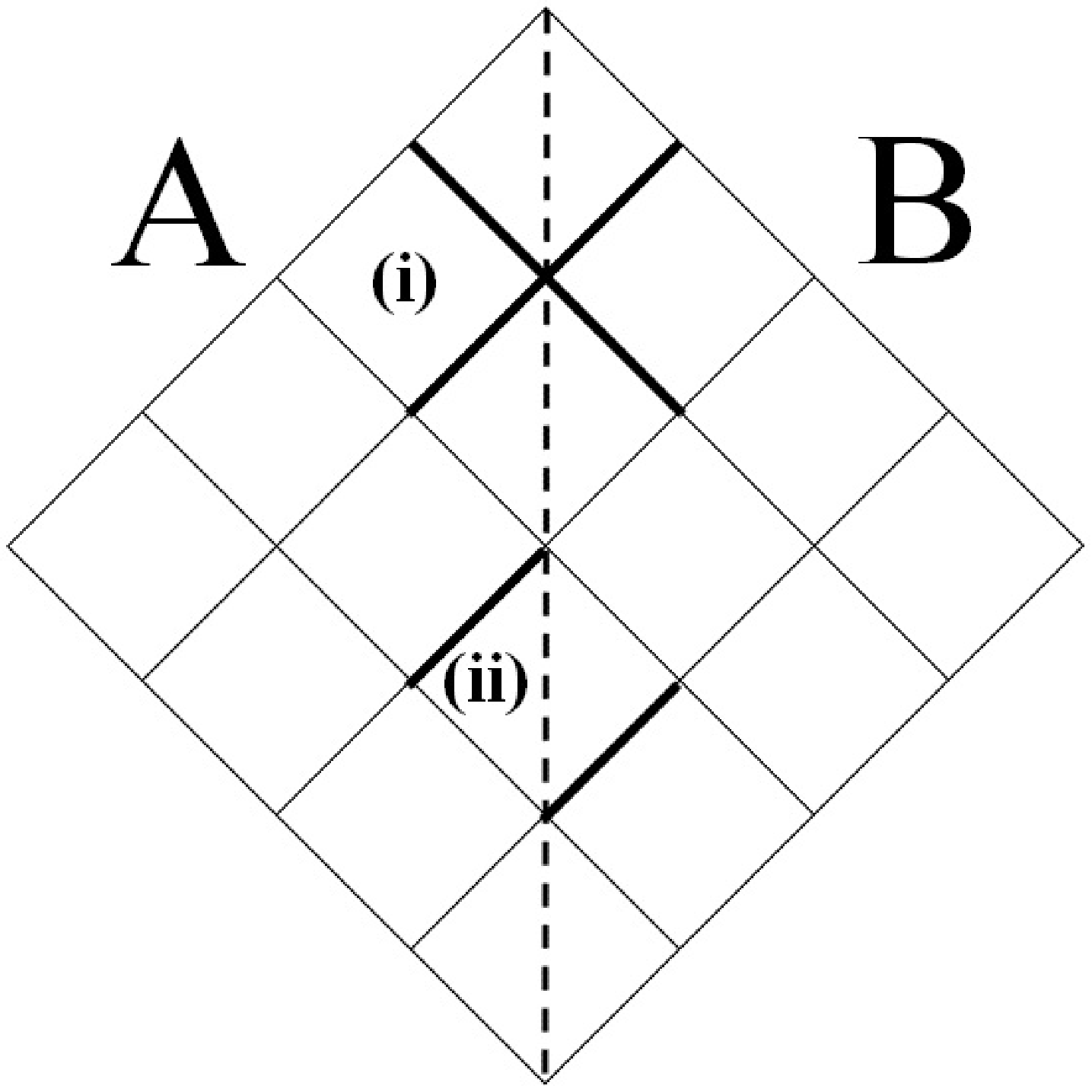}
\caption{\label{fig1} Graphs needed for the perturbative calculation
of the line entropy to $4$th order. The dashed line separates
subsystems A and B. The solid lines denote the perturbing fields in
the cluster.} \label{Fig:Graphs}
\end{center}
\end{figure}

The perturbative calculations to order $h^4$ for the Toric Code Model require just
two clusters shown in Fig.\ref{Fig:Graphs}: (i) A cluster
consisting of all four bonds incident on a boundary site between A
and B and (ii) A cluster with two bonds across a plaquette facing
each other across the boundary between A and B. The first has a
count of one per boundary site. The second has a count of 2 per
boundary site. Once grouped this way,
no subgraph subtraction is needed in $4$th order.

We now discuss the calculation of $S_n(c)$, the Renyi entropy
when only the fields on the bonds of a cluster $c$ are present.\cite{gelfand-jsp}
First the ground state wavefunction is calculated
perturbatively in the TFIM variables, with all the perturbing fields
present in the cluster. Then, for every basis state in the TFIM model,
a representative $\sigma_z$ basis state is obtained in the TCM variable.
When all the 4 perturbative fields are present at a site, the
states in the TCM must be enlarged by the operation of the
projection operator $1+A_s$ at the site. This will give us
the ground state for the cluster in the TCM variables. Once, the ground state
for the cluster is known in the TCM variables, it is straighforward
to calculate the reduced density matrix for subsystem A and hence the
entanglement entropies, $S_n(c)$.

First we consider graph (ii) from the figure. This graph has just two
bonds, one in $A$ and one in $B$. In the language of TFIM, there are
three plaquettes on which Visons can be created or destroyed.
However, since Visons are created or destroyed in pairs, there are
only 4 relevant states in the Hilbert space. These can be denoted by
the number of Visons present in the plaquette as $|1>=|0,0,0>$,
$|2>=|1,1,0>$, $|3>=|0,1,1>$ and $|4>=|1,0,1>$. These four states
can be related to the TCM in terms of $\sigma_z$ on the two bonds of
the cluster. The four states become $|1>=|1,1>$, $|2>=|-1,1>$,
$|3>=|1,-1>$ and $|4>=|-1,-1>$. In this 4-dimensional Hilbert space,
the Hamiltonian matrix is
$$H=\left(\matrix{
0 & -h & -h & 0\cr
-h & 4 & 0 & -h\cr
-h & 0 & 4 & -h\cr
0 & -h & -h & 4\cr}
\right).$$
The ground state wavefunction correct to order required for $h^4$
evaluation of Renyi entropies is
$$\psi_g=\left(\matrix{1-{h^2\over 16}-{h^4\over 512}\cr
{h\over 4}-{h^3\over 64}\cr
{h\over 4}-{h^3\over 64}\cr
{h^2\over 8} +O(h^4)\cr}
\right).$$

Let $A_i$ label the basis states for subsystem $A$ and $B_k$
label basis states for subsystem $B$. Then reduced density
matrix for subsystem $A$ has matrix elements
\begin{equation}
(\rho_A)_{A_i,A_j}=\sum_{B_k} \psi_{A_i,B_k}\psi_{A_j,B_k}
\end{equation}

In the TCM representation, there are just two states on the $A$
side, and the reduced density matrix becomes
$$\rho_A=\left(\matrix{1-{h^2\over 16}-{h^4\over 128} & {h\over 4}\cr
{h\over 4}& {h^2\over 16}+{h^4\over 128}\cr}
\right).$$

From this it follows that the two eigenvalues of the reduced density
matrix are $1-{h^4\over 256}$ and ${h^4\over 256}$. It leads to Renyi
entanglement entropy of
$$S^{(ii)}_n={1\over 1-n} [ {-n h^4\over 256} + ({h^4\over 256})^n ].$$

Now, we turn to graph (i) in figure, which has 4 bonds all sharing a
boundary site between $A$ and $B$. There are $4$ plaquettes on which
Visons can be created or destroyed. But, because they are created or
destroyed in pairs, there number is conserved modulo $2$. Thus,
there are $8$ relevant basis states. We can order the plaquettes in
a clock-wise manner. Then the $8$ states are: $|1>=|0,0,0,0>$,
$|2>=|1,1,0,0>$, $|3>=|0,1,1,0>$, $|4>=|0,0,1,1>$, $|5>=|1,0,0,1>$,
$|6>=|1,1,1,1>$, $|7>=|1,0,1,0>$ and $|8>=|0,1,0,1>$. The
$8$-dimensional Hamiltonian matrix is given by
$$H=\left(\matrix{0 & -h & -h & -h & -h & 0 & 0 & 0\cr
-h & 4 & 0 & 0 & 0 & -h & -h & -h\cr
-h & 0 & 4 & 0 & 0 & -h & -h & -h\cr
-h & 0 & 0 & 4 & 0 & -h & -h & -h\cr
-h & 0 & 0 & 0 & 4 & -h & -h & -h\cr
0 & -h & -h & -h & -h & 8 & 0 & 0\cr
0 & -h & -h & -h & -h & 0 & 4 & 0\cr
0 & -h & -h & -h & -h & 0 & 0 & 4\cr}
\right).$$
It leads to ground state wave function needed for 4th order
calculation of Renyi entropies of:
$$\psi_g=\left(\matrix{
1-{h^2\over 8}-{9 h^4\over 64}\cr
{h\over 4}+{h^3\over 16}\cr
{h\over 4}+{h^3\over 16}\cr
{h\over 4}+{h^3\over 16}\cr
{h\over 4}+{h^3\over 16}\cr
{h^2\over 8} +O(h^4)\cr
{h^2\over 4} +O(h^4)\cr
{h^2\over 4} +O(h^4)\cr}
\right).$$
Transforming to the TCM variables $\sigma_z$, there are 4 states on
the $A$ side, and remembering that each TFIM state is a
superposition of two TCM basis states after applying the
projection operator $(1+A_s$), the $4\times 4$ reduced
density matrix becomes:
$$2\rho_A=\left(\matrix{
1-{h^2\over 8}-{9h^4\over 64} & {h\over 4} +{3h^3\over
16} &
{h\over 4}+{3h^3\over 16}&{5 h^2\over 8}\cr
{h\over 4} +{3h^3\over 16} &{h^2\over 8} +{9 h^4\over 64}&
{h^2\over 4}& {h\over 4}+{3h^3\over 16}\cr
{h\over 4} +{3h^3\over 16} & {h^2\over 4} &{h^2\over 8} +{9
h^4\over 64}&
{h\over 4}+{3h^3\over 16}\cr
{5 h^2\over 8}& {h\over 4} +{3h^3\over 16} &
{h\over 4}+{3h^3\over 16}&1 -{h^2\over 8} -{9h^4\over 64}\cr}
\right).$$
From this it follows that the eigenvalues of $2\rho_A$
are $1-{3h^2\over 4}-{9h^4\over 64}$, $1+{3h^2\over
4}+{7h^4\over 64}$, ${h^4\over 64}$ and ${h^4\over 64}$.
These lead to Renyi entanglement entropies:
$$S^{(i)}_n = \ln{2} -{9n\over 32}h^4 +{1\over 1-n}[(h^4/64)^n - n(h^4/64)].$$
Note that this graph also gives the full line entropy of the
unperturbed TCM, which is $\ln{2}$ for every site that is shared
between subsystems A and B. For the unperturbed model, it is easily
shown that any larger graph will give zero upon subgraph subtraction
as no additional degeneracy arises on the line.

\begin{figure}
\centerline{\includegraphics[height=2.6in,width=3.4in]{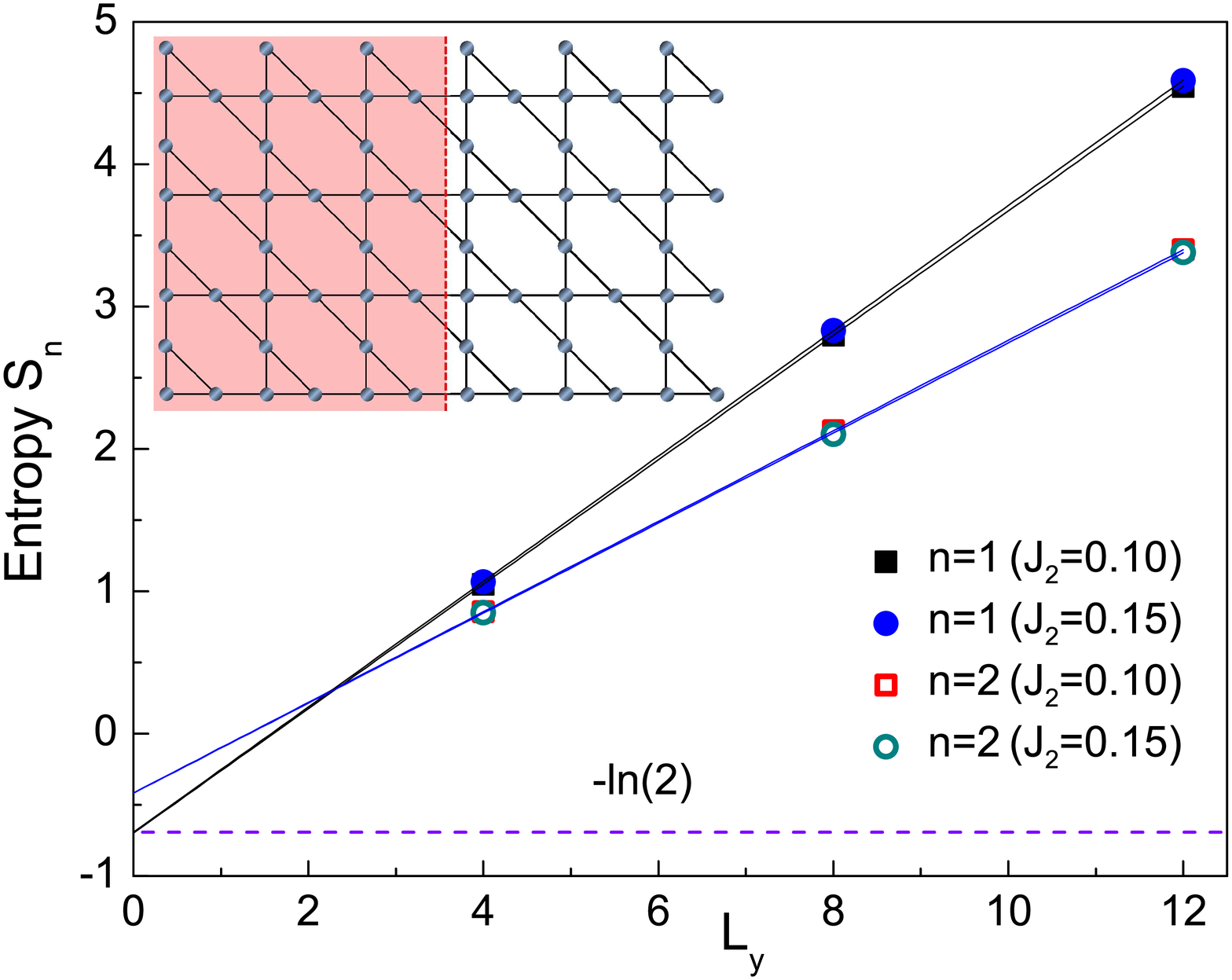}}%
\caption{The entanglement entropy $S_n(L_y)$ for the Kagome
$J_1$-$J_2$ model in Eq.(\ref{EqS:KagomeJ1J2Model}), with $L_y=4\sim
12$ at $L_x=\infty$. By fitting $S_n(L_y)=\alpha_n L_y -\gamma_n$,
we get $\gamma_1=0.698(8)$ at $J_2=0.10$, and $\gamma_1=0.694(6)$ at
$J_2=0.15$, while $\gamma_2\approx 0.42(1)$ at both $J_2=0.10$ and
$0.15$. Here $n$ is the entropy index. Inset: Kagome lattice with
$L_x=12$ and $L_y=8$.} \label{FigS:KagomeHeisenbergModelSN}
\end{figure}

\subsection{III. Kagome Heisenberg Model}%
The spin-$1/2$ Heisenberg model on the Kagome lattice, with both
first- and second-neighbor interactions, is given by%
\begin{eqnarray}
H &=&J_1\sum_{\langle ij\rangle}\textbf{S}_i\cdot
\textbf{S}_j+J_2\sum_{\langle\langle
ij\rangle\rangle}\textbf{S}_i\cdot \textbf{S}_j.
\label{EqS:KagomeJ1J2Model}
\end{eqnarray}
Here $\textbf{S}_i$ is the spin operator on site $i$, and $\langle
ij\rangle$ ($\langle\langle ij\rangle\rangle$) denotes the nearest
neighbors (next nearest neighbors). In the numerical simulation, we
set $J_1=1$ as the unit of energy. We take the kagom\'e lattice with
periodic boundary conditions along a bond direction to define a cylinder,
drawn vertically in the inset of
Fig.~\ref{FigS:KagomeHeisenbergModelSN}, and the unit of length
equal to the nearest-neighbor distance.  The results for the
entanglement entropy for $J_2=0.10$ and $0.15$ are shown in
Fig.~\ref{FigS:KagomeHeisenbergModelSN}, for which both spin-spin and
dimer-dimer correlation lengths are approximately one lattice
spacing\cite{Jiang2012TEE}.  We see that a linear fit using data for
$L_y=4\sim 12$ using Eq.(\ref{Eq:TEE}) gives $\gamma=0.698(8)$ at
$J_2=0.10$ and $\gamma=0.694(6)$ at $J_2=0.15$, both within one
percent of $\ln 2=0.693$. In contrast, the extrapolated TEE
$\gamma_n$ based on the Renyi entropy (i.e., $n\geq 2$) clearly deviates
from the expected value, although the cylinder's circumference is much
larger than the correlation length, i.e., $L_y\approx 10 \xi$. For
example, as shown in Fig.\ref{FigS:KagomeHeisenbergModelSN}, a
linear fit using Eq.(\ref{Eq:TEE}) gives us $\gamma_2\approx
0.42(1)$ at both $J_2=0.10$ and $J_2=0.15$, which differs from the
expected value by an $\approx 40\%$ error.

\end{document}